# Ultra-stable 2D/3D hybrid perovskite photovoltaic module

Giulia Grancini[1], Cristina Roldán-Carmona[1], Iwan Zimmermann[1], David Martineau[2], Stéphanie Narbey[2], Frédéric Oswald[2], Mohammad Khaja Nazeeruddin[1, *]

[1]Group for Molecular Engineering of Functional Materials, Institute of Chemical Sciences and Engineering, Ecole Polytechnique Fédérale de Lausanne, CH-1951 Sion, Switzerland.

[2] Solaronix S. A. Rue de l'Ouriette 129, 1170 Aubonne, Switzerland.

*Correspondence to: mdkhaja.nazeeruddin@epfl.ch

Hybrid perovskite solar cells, with their power conversion efficiency (PCE) exceeding 22%, have been representing a revolutionary concept for future energy power generation. Although listed among the "Top 10 Emerging Technologies of 2016", device longevity is the actual bottleneck for their real uptake in the market. Here we design an ultra-stable molecular junction of two/three dimensional (2D/3D) perovskites. It consists of a 2D $(HOOC(CH_2)_2NH_3)_2PbI_4$, anchored at the oxide substrate, that templates the growth of a highly ordered 3D $CH_3NH_3PbI_3$ perovskite stabilizing in the orthorhombic phase, even at room temperature. The unique and exceptional 2D/3D structure yields 14.6% PCE in solar cells with Spiro-OMeTAD and Au, and 12.9% PCE in hole-conductor free architecture. Aiming at the up-scaling of this technology, we realize 10x10 cm$^2$ large-area photovoltaic modules by a low-cost, fully printable, industrial-scale process delivering 11.2% PCE. We demonstrate a record stability in the PCE of 5,000 hours, setting the direction for the new generation of carbon free energy.

**Introduction**

Despite the incredible ramping up of perovskite solar cells, surpassing 22% PCE[1-4], poor perovskite solar cell longevity [5-10], mainly caused by unstability to moisture and to elevated temperature [12, 13], is at the centre of a hot debate at international research scale[12, 13]. Different possible solutions, such as the inclusion of cross-linking additives [6] or the use of barrier $NiO_x$ layers [5], have been proposed leading to a maximum PCE of 16% and 1,000 h stability. On the other hand, 2D perovskite are known for their enhanced stability due to their high water resistance[12-14]. However, their narrow absorption spectrum poses a limit to the overall device efficiency[9, 14-16]. Very recently, Ruddlesden–Popper 2D perovskite has been demonstrated to reach PCE=12% [17]. However they still suffer from degradation, reducing over 40% their PCE after running for 2,250 hours [17]. Here we develop a novel concept by engineering a molecular junction of mixed 2D/3D perovskite that we employ as active material for ultra-stable photovoltaic devices. Inspired by the concept of crystal engineering exploiting hydrogen bonds and supramolecular synthons in 2D layered perovskite [18-19], we have developed a pure 2D and a combined 2D/3D perovskite by using the ammoniumvaleric acid iodide ($HOOC(CH_2)_4NH_3I$, AVAI hereafter) as organic precursor mixed with methylammonium iodide ($CH_3NH_3I$) and the inorganic $PbI_2$ (see Methods for details). The solution is deposited on top of a mesoporous $TiO_2$ or $ZrO_2$ substrate.

**Experimental Results and Discussion**

The absorption and photoluminescence (PL) spectra of the samples are represented in Figure 1a, b, respectively. For the 2D perovskite, a clear band edge appears at 450 nm along with a remarkable peak at 425 nm, suggesting the presence of a stable excitonic transition[18], while it emits at 453 nm. In a similar fashion, to engineer the 2D/3D molecular junction, we mix the

(AVAI:PbI$_2$) with (CH$_3$NH$_3$I:PbI$_2$) at different molar ratios (0 – 3 – 10 – 20 – 50 – 100 %) in the total volume (see Methods for details). This solution is infiltrated in the mesoporous oxide scaffold upon a single step deposition followed by a slow drying-process, allowing the reorganization of the components in the film before solidification. The absorption edge of the mixed perovskites (Figure 1a) appears at 760 nm, indicating that a 3D CH$_3$NH$_3$PbI$_3$ perovskite phase is formed. In addition, an enhanced peak at 430 nm is observed, that gains intensity increasing the percentage of the (AVAI:PbI$_2$) (see also Figure S1) is observed. Figure 1b shows, for comparison, the PL of the mixed 2D/3D 3%(AVAI:PbI$_2$) perovskite: a weak peak at 453 nm appears, spectrally matching with the (HOOC(CH$_2$)$_2$NH$_3$)$_2$PbI$_4$ emission (see also Figure S2). These observations demonstrate to the co-formation of a pure 2D phase within the mixture. In addition, since the 2D emission is observed only upon excitation from the oxide side, the 2D phase must be retained at the interface with the oxide surface due to the favourable anchoring of the carboxylic acid group and the oxide surface. The formed junction is depicted in the cartoon in Figure1d.

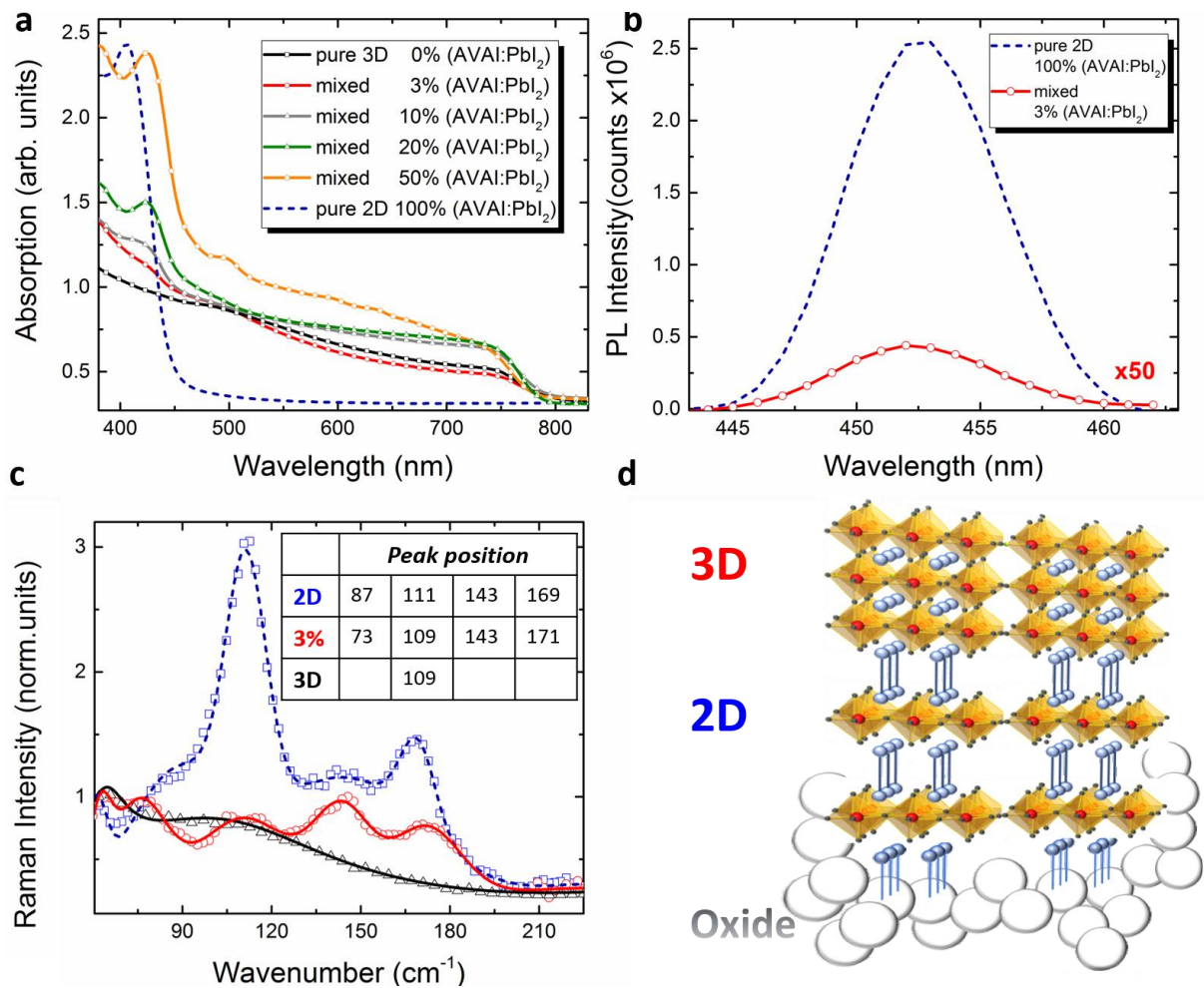

**Figure 1. Optical and Raman characterization of 2D and mixed 2D/3D perovskite. a.** Absorption spectra of the mixed perovskite with increasing % of (AVAI:PbI$_2$), as indicated in the legend, compared to 3D CH$_3$NH$_3$PbI$_3$ and 2D (HOOC(CH$_2$)$_2$NH$_3$)$_2$PbI$_4$ infiltrated into mesoporous TiO$_2$. **b.** PL spectra of the 2D and mixed 3%(AVAI:PbI$_2$) perovskite. **c.** Raman spectra for CH$_3$NH$_3$PbI$_3$, 2D and 3% (AVAI:PbI$_2$) samples. Solid lined represent the fit from multi-gaussian peaks fitting procedure. Table with fitting results in the inset. **d.** Schematic cartoon of the 2D/3D molecular junction.

Figure 1c shows the micro-Raman spectra in the low wavenumber range of the 2D, 3D and mixed 3% (AVAI:PbI$_2$) perovskite. The (HOOC(CH$_2$)$_2$NH$_3$)$_2$PbI$_4$ shows Raman peaks at 87 cm$^{-1}$,

111 cm$^{-1}$ and 169 cm$^{-1}$ associated to the vibration modes of PbI$_2$ layered crystal [20, 21]. On the other side, the peak at 143 cm$^{-1}$ can be assigned to the vibrational mode of the organic cation. This mode results from the intercalation, in the 2D perovskite, of the organic moiety that disrupts the stacking sequence continuity of the PbI$_2$ layer, similarly to what observed for PbI$_2$ crystal intercalated with pyridine or polyaniline [21, 22]. The presence of the same peaks is also observed in the mixed 2D/3D perovskite, confirming the presence of the 2D phase. However, these peaks stand on broader features that are characteristic of the CH$_3$NH$_3$PbI$_3$ [20]. Figure 2a shows the X-Ray-Diffraction (XRD) pattern of the 2D/3D junction compared with the 2D perovskite. The (HOOC(CH$_2$)$_2$NH$_3$)$_2$PbI$_4$ perovskite shown an XRD pattern characteristic of the crystal in the *Pbca* centric space group, in agreement with what previously observed (see also Figure S3a)[11, 17].

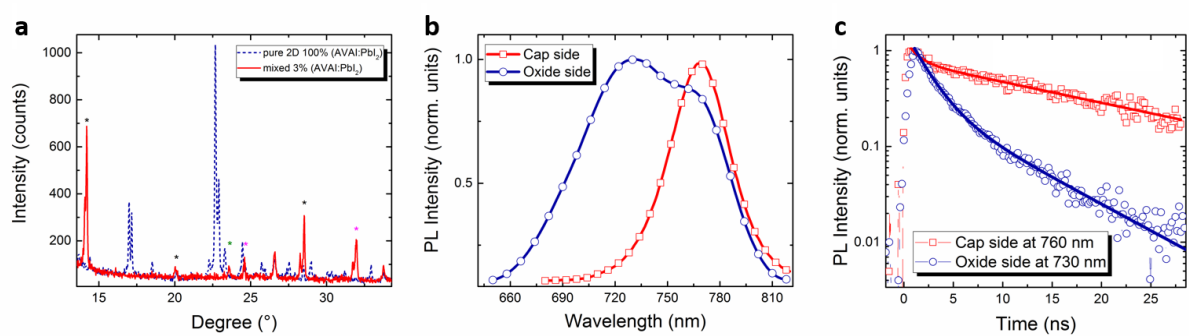

**Figure 2. XRD and Photoluminescence analysis of the 2D and mixed 2D/3D perovskite. a.** XRD pattern of the 2D and the 3% (AVAI:PbI$_2$) deposited on the TiO$_2$ scaffold. **b.** Normalized PL spectra, excitation at 600 nm, for the mixed 3%(AVAI:PbI$_2$) exciting from the capping side and from the oxide side. Since the light penetration depth is < 100 nm at 600 nm, excitation of the perovskite film from the oxide side (scaffold thickness ∼ 1μm) interrogates the perovskite nano-crystallites grown within the scaffold while excitation from the capping layer side (∼ 1μm thick) probe the intrinsic properties of the capping layer. **c.** PL dynamics of the bulk perovskite (exciting from the capping layer) at 760 nm and of the interface with the oxide at 730 nm of the 3%(AVAI:PbI$_2$). Insulating ZrO$_2$ mesoporous substrate is used to exclude any additional injection process.

The XRD pattern of the mixed 2D/3D corresponds to the tetragonal phase of the CH$_3$NH$_3$PbI$_3$[11], marked with black asterisks, along with additional reflections (marked with green and pink asterisks). The latest, at 23.2° and 33.9°, are the marker of a preferential growth along the oriented c-axis direction[11], typical of the 3D CH$_3$NH$_3$PbI$_3$ at the low-temperature orthorhombic phase[23]. Although this has been already observed by few of us [11], the reason behind has been so far completely ignored. On the other side, the diffraction peak at 23.5° matches with the one of (HOOC(CH$_2$)$_2$NH$_3$)$_2$PbI$_4$ , although it is very weak when using less than 50% (AVAI:PbI$_2$) (see Figure S3b). These measurements confirm the co-existence of the 2D and 3D phases but, in addition, reveal that the CH$_3$NH$_3$PbI$_3$ partially assembly in the orthorhombic phase stable at room temperature. Figure 2b compares the PL spectra of the perovskite crystal confined in the mesoporous oxide with the behaviour of the bulk crystal of the capping layer on top by selectively exciting from the two different sides (see Figure 2b). The PL from the capping layer reveals peaks at around 760 nm, while, on the other side, PL from the mesoporous oxide reveals an additional peak at 730 nm. This blue-shifted PL, never observed so far for the CH$_3$NH$_3$PbI$_3$PbI$_3$ at room temperature, is related to a 3D CH$_3$NH$_3$PbI$_3$ with a wider band gap of 1.69 eV, typical of the orthorhombic phase appearing below 160 K [24-26]. Interestingly, by getting rid of the capping layer only the peak at 730 nm appears (Figure S4). Figure 2c compares the PL dynamics of the two different bands exciting both sides, respectively. At 760 nm the PL shows a long-living decay (extending over our temporal window) due to electron-hole recombination[27], while at 730 nm fast component with a time constant of $\tau$=2ns dominates (see Table S2). Note that we intentionally use the insulating $ZrO_2$ substrate to avoid any additional processes. However, a similar behaviour is observed also on the $TiO_2$ substrate, confirming that it is intrinsically related to the mixed 2D/3D structure (see Figures S5, S6). The 2ns decay is already observed from $CH_3NH_3PbI_3$ at low temperature [24, 25], due to a contribution of exciton recombination that is stabilized in the orthorhombic phase [28]. The combined XRD and PL analysis demonstrate the unique role of the 2D perovskite, anchored at the interface with the oxide nanoparticle network, in templating and stabilizing the growth of the $CH_3NH_3PbI_3$ in the orthorhombic phase. This induces a vertical segregation of the material, where the wider band-gap $CH_3NH_3PbI_3$ is formed within the oxide pores, while the tetragonal $CH_3NH_3PbI_3$ perovskite forms on top. Motivated by these observations, we have fabricate solar cells using this material as active layer and mesoporous- $TiO_2$ and spiro-OMeTAD as charge transport layers. Figure 3a shows the J-V characteristic of the mesoporous 2D/3D perovskite with 3%(AVAI:$PbI_2$) as the optimized composition (see also Figure S7). The device shows a record PCE of 14.6% (see device statistics in Figure S8) that is maintained up to 60% of the initial value after 300 hours illumination (see Figure 3b), indicating a reasonably good device stability. However, recent studies have revealed the use of Spiro OMeTAD and Au electrode as not optimal option for cheap and long-term stable solar cells [29]. Therefore we have also fabricated 2D/3D perovskite within hole-conductor free solar cells architecture [11].

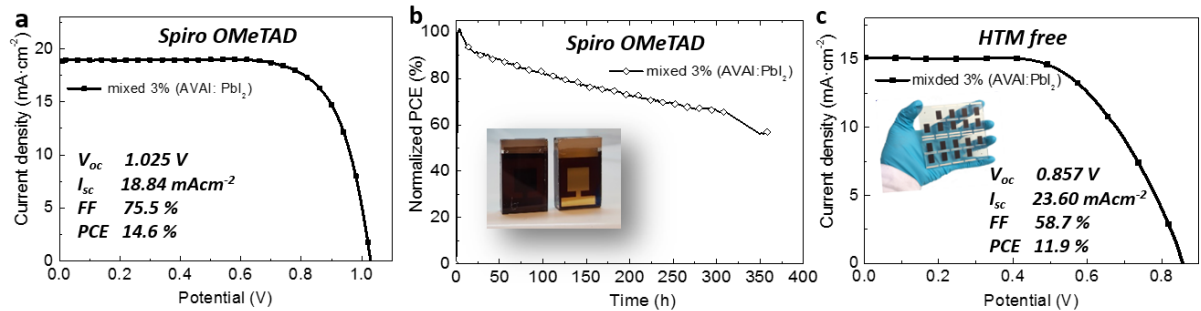

**Figure 3. Solar Cell Characterization.** Current density – Voltage (J-V) curve obtained for a typical mixed 2D/3D perovskite with 3% (AVAI:PbI$_2$) measured under AM 1.5G illumination **a.** using Spiro-OMeTAD as hole transport material and Au as the electrode; **b.** stability curve measured at maximum power point under AM 1.5G illumination and stabilize temperatura of 45° for the device containing Spiro-OMeTAD and Au layers. **c.** Solar cell using the hole conductor free configuration suppressing the HTM layer and using carbon as the electrode. Insets: Pictures of the devices fabricated with Au and carbon electrodes.

The solar cell delivers a maximum 12.9% PCE (see also Table S1). The J-V characteristic of 1cm$^2$ solar cell is shown in Figure 3c. Since this configuration holds the promise to be at present the cheapest and the most attractive solution among the perovskite photovoltaic architectures, we have developed a 10x10 cm$^2$ large area solar modules in a fully printable hole-conductor free architecture. Figure 4a shows the J-V curve of the module, delivering a champion PCE=11.2% (see Table S1). We have tested the modules under different conditions including AM 1.5 G solar illumination at 1000 W/m$^2$ and cycling of temperature up to 90°. The results, in Figure 4b, c show extraordinary long-term stability of at least 4,752 h and excellent response at elevated temperature. Such long stability is at present the highest value obtained for perovskite photovoltaics (see Figure S9).

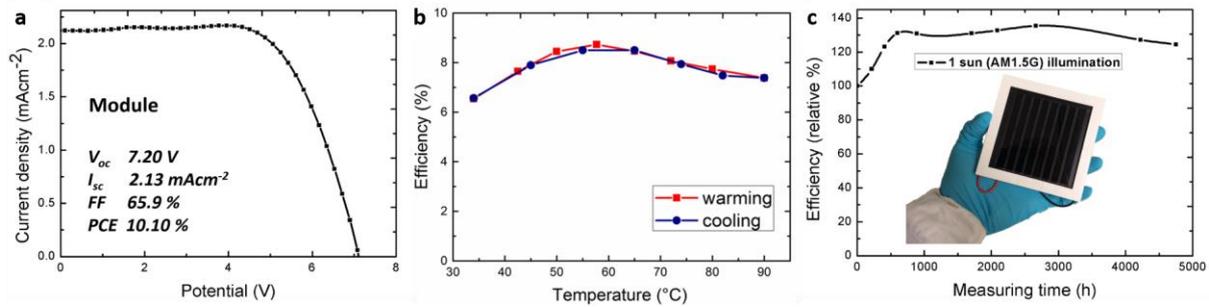

**Figure 4. Large area Photovoltaic Module Characterization.** **a.** IV characteristics of the module. **b.** response of the module efficiency versus temperature stress upon warming (red) and cooling (blue) measured at 0.75 sun. **c.** Typical module stability test under 1 sun AM 1.5 G conditions for 4,752 h. In the inset picture of the module.

**Conclusions**

In conclusion, we molecularly engineer a built-in 2D/3D perovskite which grows forming a bottom-up phase-segregated structure. The 2D perovskite acts as a protective barrier and templates the growth of a highly ordered and stable $CH_3NH_3PbI_3$ perovskite on top. Used as active film in hole-conductor free, fully printable, 10x10 cm$^2$ large solar module, we demonstrate 11.2% PCE and a record stability of at least 5000 h, surpassing with a gigantic step the results obtained so far. In addition, the fundamental insights highlighted in this study, i.e. the growth of $CH_3NH_3PbI_3$ in two different phases that co-exist at room temperature, will have enormous consequences on the physics behind this fascinating material (i.e. the co-presence of exciton –in the orthorhombic phase and charge population– in the tetragonal phase within the same nanoscale material). This results will open up new research directions aiming at better exploring the potentiality of this hybrid 2D/3D structure over many disciplines from material science and chemistry to solid state physics.

# References

1. Saliba, M. *et al*. A molecularly engineered hole-transporting material for efficient perovskite solar cells. *Nat. Energy* **1**, 15017 (2016).

2. Yang, W. S. *et al*. High-performance photovoltaic perovskite layers fabricated through intramolecular exchange. *Science* **348**, 1234–1237 (2015).

3. "National Renewable Energy Laboratory Best Research-Cell Efficiencies," can be found under http://www.nrel.gov/ncpv/images/efficiency_chart.jpg.

4.http://www3.weforum.org/docs/GAC16_Top10_Emerging_Technologies_2016_report.pdf

5. You, J. *et al*. Improved air stability of perovskite solar cells via solution-processed metal oxide transport layers. *Nat. Nanotechnol.* **11**, 75–81 (2016).

6. Li, X. *et al*. Improved performance and stability of perovskite solar cells by crystal crosslinking with alkylphosphonic acid ω-ammonium chlorides. *Nat. Chem.* **7**, 703–711(2015).

7. Kaltenbrunner, M. *et al*. Flexible high power-per-weight perovskite solar cells with chromium oxide–metal contacts for improved stability in air. *Nat. Mater.* **14**, 1032–1039 (2015).

8. Han, Y. *et al*. Degradation observations of encapsulated planar $CH_3NH_3PbI_3$ perovskite solar cells at high temperatures and humidity. *J. Mater. Chem. A Mater. Energy Sustain.* **3**, 8139–8147 (2015).

9. Smith, I. C., Hoke, E. T., Solis-Ibarra, D., McGehee, M. D. & Karunadasa, H. I. A layered hybrid perovskite solar-cell absorber with enhanced moisture stability. *Angew. Chem. Int. Ed.* **53**, 11232–11235 (2014).